# Quantifying stretching and rearrangement in epithelial sheet migration

**Rachel M. Lee[1], Douglas H. Kelley[2, 3], Kerstin N. Nordstrom[1], Nicholas T. Ouellette[3], Wolfgang Losert[1]**

[1] Department of Physics and Institute for Research in Electronics and Applied Physics, University of Maryland, College Park, MD, USA

[2] Department of Materials Science & Engineering, Massachusetts Institute of Technology, Cambridge, MA, USA

[3] Department of Mechanical Engineering & Materials Science, Yale University, New Haven, CT 06520, USA

E-mail: rmlee@umd.edu

**Abstract**

Although understanding the collective migration of cells, such as that seen in epithelial sheets, is essential for understanding diseases such as metastatic cancer, this motion is not yet as well characterized as individual cell migration. Here we adapt quantitative metrics used to characterize the flow and deformation of soft matter to contrast different types of motion within a migrating sheet of cells. Using a Finite-Time Lyapunov Exponent (FTLE) analysis, we find that - in spite of large fluctuations - the flow field of an epithelial cell sheet is not chaotic. Stretching of a sheet of cells (i.e., positive FTLE) is localized at the leading edge of migration and increases when the cells are more highly stimulated. By decomposing the motion of the cells into affine and non-affine components using the metric $D^2_{min}$, we quantify local plastic rearrangements and describe the motion of a group of cells in a novel way. We find an increase in plastic rearrangements with increasing cell densities, whereas inanimate systems tend to exhibit less non-affine rearrangements with increasing density.

**1. Introduction**

Collective motion at the cellular level plays an important role in many biological processes. From development [1,2] to wound healing [3–5], cells must cooperate to create complex structures. Faulty regulation of collective behavior can also disrupt development or lead to diseases such as metastatic cancer [6–9]. Despite the abundance of collective migration in biological systems, it is not yet as well characterized as individual cell movement.

One way to begin to understand how collective behavior emerges from individual motion is to make analogies to non-biological physical systems. Migrating cells share some features with physical systems; the migration of a group of cells calls to mind images of flowing fluids or sheared solids. Yet cells' active properties often allow the group to move in ways that are distinct from inanimate physical systems. Current research has begun to look for similarities and differences between active cells and passive physical systems by exploring the stresses the cell experience [6,10–14] and the cell-cell/cell-matrix adhesions that hold sheets of epithelial cells together [5,9,15–18]. The changes seen in migration as cells reach different densities [19,20] have also motivated comparisons to kinetic phase transitions [21,22]. Continued application of physical tools to groups of cells will undoubtedly reveal interesting features of migration – both those that are shared with inanimate systems and those that are unique to active, living matter.

One particularly interesting physical feature of collectively migrating cells is their velocity field. Velocity fields have long been studied in physical systems such as fluid flows or granular systems. By applying techniques developed to study physical velocity fields to the information present in live cell imaging, we have developed tools to quantify the speed, density, and rearrangement of epithelial sheets during collective cell migration.

One way to extract velocity information from live cell images is particle image velocimetry (PIV). This technique, originally developed to study fluid flows, has begun to see use in biological systems (see for example the work by Petitjean et al. [23]). PIV estimates the average flow field from one image to the next by correlating the two images, but, unlike cell tracking, it does not provide information about individual cell motion. PIV analysis is easy to carry out on phase contrast images and provides a smoothed flow field that describes the overall motion of the migrating sheet. Since PIV assess the average motion of all features visible in an image, it takes into account all components in the cell. A second common technique for extracting velocity information is to track each cell's nucleus. Although PIV is able to be applied to both fluorescent and phase contrast images, tracking generally relies on fluorescent images. However, techniques for tracking objects in phase contrast images are steadily developing [24].

To determine the stretching rate of the flow field determined using PIV analysis, we use Finite-Time Lyapunov Exponents (FTLE). The FTLE is a measure of how much two points initially near each other in the flow field would separate after a given time interval [25–27]. Similar to the traditional Lyapunov Exponent, which quantifies separation on an asymptotic time scale, the average FTLE reflects how chaotic the flow is within a finite time (in our case, we choose a time scale over which the cells move roughly one cell length). Measuring an FTLE *field* adds spatial information about which regions are more or less chaotic. This allows us to assess whether the chaotic features of the flow field are correlated with a cell's distance from the leading migration edge or other spatially localized features.

The tracks give information about individual cells; by combining PIV data with individual cell tracks we can consider the motion and flow of the migrating epithelial sheet. To parameterize an individual cell's deviation from the motion predicted by the average flow field, we use the previously defined parameter $D^2_{min}$, which quantifies the non-affine deformation around a given cell [28–32]. A cell with a high value of $D^2_{min}$ moves in a distinct manner compared to what is predicted by the smooth local flow field. Regions in the sheet with high $D^2_{min}$ can be considered regions of high rearrangement.

To illustrate the use of these two metrics, we analyze a migrating two-dimensional culture of epithelial cells. We find that statistics such as average speed vary from experiment to experiment, even under the same conditions. However, by considering our metrics for rearrangement and stretching, we are able to quantify trends that hold across different experiments. We find that despite the radial migration of our monolayer, most of the stretching of the sheet is limited to a small region near the leading migration edge. We also find that rearrangements within our cell sheet increase as the local cell number density increases, a surprising trend that could be used to isolate active properties of the migrating cells. These results suggest that $D^2_{min}$ and FTLE data will be useful in distinguishing and classifying how cells migrate collectively in many different systems under varying conditions, making them a promising tool for future cancer studies.

## 2. Materials and Methods

### *2.1 Cell Culture and Microscopy*

MCF10A cells were cultured in DMEM/F12 media supplemented with 5% horse serum, 10 μg/ml insulin (Invitrogen), 10 ng/ml EGF (Peprotech, Rocky Hill, NJ), 0.5 μg/ml hydrocortisone, and 100 ng/ml cholera toxin (both Sigma, St. Louis, MO). The cells were kept in a humidified atmosphere at 37 °C and 5% $CO_2$. For the migration assay, cells were plated in 12 well glass bottomed plates coated with collagen IV (10μg/ml for approximately three days at 4 °C) and allowed to culture overnight in DMEM/F12 containing 1% horse serum (unless otherwise noted to be 5% horse serum). Figure 1 provides an illustration of the plating technique. Cells were then imaged for 24 hours on an incubator microscope kept at 37 °C and 5% $CO_2$ (Zeiss Observer.Z1, Zeiss, Goettingen, Germany). Phase contrast images of the cells were acquired at 2 minute intervals using a 10x objective.

### *2.2 Image Analysis*

Particle Image Velocimetry (PIV) analysis is performed using the MatPIV MATLAB toolbox (J. Kristian Sveen: http://folk.uio.no/jks/matpiv/, GNU general public license). We use multiple iterations of interrogation window sizes, starting with two iterations of 64 x 64 pixel windows (41.6 μm x 41.6 μm) and finishing with two iterations of 32 x 32 pixel windows (20.8 μm x 20.8 μm). At each interrogation, 50% overlap is used and the windows are cross-correlated. After the final iteration, outliers are detected using a signal-to-noise filter. Vectors with a signal-to-noise ratio less than 1.3 are replaced by linearly interpolated values. In the case of movies involving a leading edge of the cell sheet, the edge of the sheet is segmented and the detected edge is used to trim the PIV field to remove the zero velocity contribution of the empty space from any further metrics.

We track individual nucleoli using an algorithm adapted from hydrodynamic turbulence studies, implemented in Matlab and available online [33]. We begin by calculating each data set's mean image, which we take as the background. Wherever an individual image is sufficiently darker than the background, over a region of appropriate size, the algorithm identifies a dot (possible nucleolus), taking for its position the intensity-weighted centroid of the region. These dots are tracked with a predictive three-frame best-estimate algorithm [34], yielding the location of each dot over a range of frames. Because we are interested in the motion of cells, we then seek groups of dots that remain within a set distance of each other and replace those individuals with their centroid, approximating the motion of a cell's nucleus. We typically choose parameters that identify many dots, since we want to include as many real cells as possible and false positives produce short tracks with little statistical weight. In one example frame we hand-counted 674 dots in 372 cells, whereas our automated algorithm identified 817 dots, reducing that count to 504 after grouping. Automated tracking of dots in phase-contrast images gives us access to the individual motion of many more living cells than would be practical when counting by hand.

### *2.3 Field of View Reconstruction*

For each of the circular monolayers of cells, four fields of view were recorded. In two of the image sequences, the field of view is entirely filled with cells and will be referred to as an image of the "bulk." In the other two image sequences, referred to as the "edge" movies, the field of view includes the leading edge of the cells and the empty space in front of the monolayer. Our initial analysis focuses on the comparison of behaviors between these two sets of movies with no further spatial distinction.

To probe the differences seen between edge and bulk migration behavior in more detail, we perform a segmentation of the leading edge to obtain an outline of the monolayer over time. This segmented edge is then fit to a circle, allowing us to find an effective radius and center of the monolayer as a function of time. Each cell's position is converted into the radial distance from the center of the monolayer. The radial distance of a given cell from the center, *r*, normalized by the total radius of the monolayer, *R*, allows us to compare cell motion to location within the sheet.

*2.4 Finite Time Lyapunov Exponents*

Working from PIV velocity fields, we calculate finite-time Lyapunov exponents (FTLEs), as shown in Figure 4. FTLEs are calculated by numerically advecting "virtual" tracers through the measured, time-varying flow field. We initially seed tracers on the PIV grid but subsequently allow them to move off of the grid. We follow the tracer motion for a deformation time of 80 minutes. Because FTLEs asymptotically approach Lyapunov exponents, our results are not sensitive to the deformation time if it is reasonably long [25]. At the end of the deformation, we calculate the logarithm of the largest eigenvalue of the Cauchy-Green strain tensor for each cluster of four tracers that were initially neighbors; this is the local FTLE value. By measuring one FTLE value for each point on the PIV grid, we obtain the full FTLE field [26].

*2.5 Non-Affine Motion*

To quantify rearrangements within the sheet, we look at non-affine motion. Specifically, we measure relative motion at the single cell level - for each cell we look at the relative displacements of its neighbors within a 40 μm radius. We then calculate the quantity $D^2_{min,i}$, [28,29] which quantifies the non-affine deformation of the *N* neighbors (indexed by *j*) of cell *i*:

$$D^2_{min,i} = \min\left\{ \frac{1}{N} \sum_j \left[ \Delta\vec{d}_{ij} - E_i \vec{d}_{ij} \right]^2 \right\},$$

where the vector $\vec{d}_{ij}$ is the relative initial position of cells *i* and *j*, $\Delta\vec{d}_{ij}$ is their relative displacement, and the minimization is over all potential strain tensors $E_i$. In other words, we calculate an affine deformation field ($E_i$) that best describes the actual cell motion. If the motion around cell *i* were perfectly affine, $D^2_{min,i}$ would be zero; $D^2_{min,i}$ is larger for greater departures from purely affine motion. $D^2_{min,i}$ thus gives us the relative strength of non-affine motion, or plastic rearrangements, in the neighborhood of cell *i*.

## 3. Results

### *3.1 Velocities*

Cells were added to collagen-coated coverslips in such a way as to form a roughly circular monolayer of cells (see Figure 1 for a schematic of the assay). Time-lapse images were acquired at multiple locations of the dot: some within the bulk, and others at the leading edge. Both PIV and tracking of cell nuclei show significantly higher speeds for cells near the leading edge of the monolayer, as shown in Figure 2(a). That the tracking shows a slight increase in mean speed compared to PIV is to be expected from the properties of the methods. PIV gives a smoothed velocity field that will suppress the contribution of isolated fast motion to the overall mean speed. It also measures all motion between two frames, including membrane and cytoplasm fluctuations in addition to the nuclear motion; this randomly directed motion may lead to smaller average speeds. Nuclear tracking, however, follows each cell's nucleus individually and retains information about isolated outliers. The two methods capture different aspects of the system and thus provide slightly different information.

In the case of PIV, dividing cells and the few visible dead cells are included in the velocity information. In contrast, tracking does not include dead cells and is only able to measure dividing cells while the tracked features (nucleoli) are visible as during the very beginning and end of the division process. In a subset of movies, cell divisions were tracked and showed no correlation with changes in either speed metric (data not shown).

Despite the agreement between all experiments on the overall trend between edge and bulk, the experiments still exhibit a wide variety in speeds from day to day. As shown in Figure 2(b), the mean speeds from one experiment are significantly higher. Although all three experiments show a significant increase in that day's edge speed compared to the same day's bulk speed, the bulk cells from experiment

C move as quickly as the edge cells from experiments A and B. The variation in speed could result from sensitivity to initial conditions such as small differences in the coating of a microscope slide. The authors' experience with migration assays suggests that small variations in these parameters and variation in speeds are frequently observed. The results shown in Figure2(b) emphasize that it is important to be careful when comparing experiments performed on different days, and suggest the need for additional quantitative metrics when comparing the dynamics of epithelial sheet migration.

*3.2 Finite Time Lyapunov Exponents*

Figure 4 shows an example snapshot of FTLE values. We find that the spatial average of FTLEs is negative in all cases, and that it approaches zero as the deformation time *T* increases. Negative FTLEs imply that the flow is not chaotic, and that neighboring regions tend to separate only slowly as time passes. Slow separation is consistent with the fact that the majority of our observations show bulk cells surrounded by many neighbors, so their relative motion is highly constrained. However, the resolution of our data prevents the FTLE calculations from accounting for length scales smaller than the PIV grid; the flow may be chaotic at these small scales.

By fixing the deformation time at *T*=80 min we can consider FTLE fields in more detail, as shown in Figure 5. Distributions of FTLE magnitude show that local values can vary substantially from the mean, occasionally being greater than zero. Positive FTLEs are more common in data sets recorded at the advancing edge of a cell culture than in the bulk, as indicated both by distributions and by snapshots—the edge experiences large stretching. This observation is borne out in FTLE statistics conditioned on the normalized radius *r/R* (as shown in Figure 6(c)).

Many simple flows are chaotic, regardless of the flow speed [25, 26], so the lack of chaotic motion is not solely due to the relatively slow speed of the migrating sheet. In fact, the chaos measured in the sheet is changed with modifications to the migration environment, as shown by the 5% serum data in Figure 5. Serum contains many compounds, including growth factors and other stimulants, which can change the sheet's mode of migration. The change in migration activity is seen in the shift towards more chaotic motion in the bulk FTLE distributions (Figure 5(a)) and even more dramatically in the distribution of edge FTLE values (Figure 5(b)). Although in all cases the mean FTLE remains negative and thus not chaotic, in the spatial information provided by the FTLE field we find instances of positive values, i.e., chaotic motion. Although the speed of the sheet was moderately increased by the addition of serum (Figure 6(b)), the relative quantity of positive values was many fold increased in migration at higher concentrations of serum (Figure 5(c)). The more active motion provided by the stimulation at 5% serum is reflected in the increased instances of chaotic motion.

### *3.3 Non-Affine Motion*

$D^2_{min}$ is a spatially and temporally heterogeneous quantity: “hot spots” of plastic rearrangements appear and die out after the rearrangement is complete. Figure 3(a) shows these hot spots in one frame of a bulk movie. We look at the differences in $D^2_{min}$ between the edge and bulk in Figure 3(b). $D^2_{min}$ is normalized by the mean-squared displacement, so that the magnitude of $D^2_{min}$ is not dependent on the macroscopic flow speed. Clearly, cells in the bulk undergo more plastic rearrangements than on the edge, and this trend holds for different experiments.

To investigate this result further, we also look at the probability density of $D^2_{min}$ values for several different experiments, shown in Figure 3(c). $D^2_{min}$ has units of area, so we have translated the *x*-axis into the corresponding length. We see two characteristic peaks in the distribution, one near 2 μm and one near 10 μm. The peak at longer lengths grows for the bulk datasets, suggesting that the primary mode of rearrangements changes in the bulk. The length scale associated with this peak is comparable to the cell size, suggesting that these are actual position-switching deformations.

We also stitch together data from bulk and edge to tease out the dependence of $D^2_{min}$ on distance from the center, as shown in Figure 6(d). The basic trend shows that non-affine motion increases into the bulk and eventually plateaus. Since the cell density changes from bulk to edge, we next consider the dependence of $D^2_{min}$ on density.

*3.4 Density Effects*

After measuring consistent bulk versus edge trends in our metrics and yet seeing a clear variability between experiments, we look to the original images for variation in factors other than speed. We find apparent differences in the edge densities between experiments, prompting the comparison of our metrics to the local cell density. We calculate local cell density by counting the number of neighboring nuclei contained within a 40 μm radius of the cell nuclear location.

As seen in Figure 6(a), there is a clear distinction between the local neighbor count profiles for each experiment. These density differences are accompanied by variation in the radial speed profiles as seen in Figure 6(b). Within each day of experiments, four migration assays were performed. The density and speed trends are consistent across all four assays on a single day – speed decreases as density increases. For two of the days, a clear decrease in number density is seen near the leading edge of the migrating cells. This is accompanied by a smooth speed profile: a nearly constant speed in the center that quickly increases to faster speeds at (the lower density) edge. For the third day (Exp C in Figure 6), all four migration assays show a density profile that differs from the other two experiments. For these monolayers, a peak in density is found ~0.1*R* away from the leading edge. This different density profile is accompanied by a different speed profile: faster overall, and lacking a constant speed in the center, as seen in Figure 6(b). This set of experiments also shows a perturbation in stretching (Figure 6(c)) and rearrangements (Figure 6(d)) near the relative position of the density perturbation.

Intrigued by the differences near this perturbation, we further investigate the relationship between our metrics and the local number density. To avoid including edge effects in our investigation of number density, we restrict our analysis to bulk movies. The number of neighbors within a radius of 40 μm of each cell is used as a measure of the local of cell density. Figure 7(a) shows the relationship between cell speed and number density for three characteristic bulk movies. In all cases, although the magnitude of the speed varies from experiment to experiment, speed is a consistent function of density: cells in higher density regions are slower on average.

Although we investigated the relationship between FTLE values and local cell density, no strong trends were seen. As the FTLE analysis captures long time deformation and stretching, it is not surprising that it does not show a significant relationship to short time parameters such as the local cell density. Further, because the stretching measured by the FTLE values is relative between cells, the dependence of average speed on local cell density does not imply a FTLE dependence on density. After all, even if all the cells are moving faster, they need not be separating more quickly. The FTLE analysis is very similar to the affine motion of the cells [35], which does not have a clear relationship with density.

But we also measure the non-affine component of the motion using the parameter $D^2_{min}$. For each movie, for all frames and all cells, we determine $D^2_{min}$ as a function of the local density, as shown in Figure 7(b). $D^2_{min}$ is larger for higher densities – in a consistent manner between experiments. This relationship points to the source of the difference in $D^2_{min}$ for the edge and bulk data. The consistency between experiments is reassuring, but the implications of the trend are very intriguing. Non-affine motion increases as the density increases. This is the opposite trend one would see for a jammed system of soft particles [36]. In such jammed particulate systems, the probability of plastic deformations increases with decreasing density, as more voids are created. In contrast, as cells get denser (more jammed), they rearrange more, suggesting active, individualistic motion. This active motion is a known feature of both healthy and cancerous cells: cells in epithelial sheets actively generate traction forces on the underlying substrate [10-14] and tumor cells are known to actively infiltrate surrounding tissues [9]. Cells are able to take advantage of forces generated by their cytoskeletal structures to actively migrate using structures such as lamellipodia and filopodia [4]; these same protrusive and contractive forces may contribute to cell rearrangements within a migrating sheet.

## 4. Discussion

Using image analysis to extract velocity information from phase contrast images of cells, we were able to quantify the migration of epithelial sheets. Looking only at the simple metric of speed, we distinguished different migration activity in the center and at the edge of the radially migrating monolayer. However, we also found significant variability from experiment to experiment, due to the biological variability of our migration assay. These differences, accompanied by the evidence that smoothed velocity techniques like PIV capture elements of the system different from what is measured by commonly used nuclear tracking, provide motivation for precise quantitative metrics of the cell motion.

As a suggestion for one such metric, we presented FTLE fields for the migrating sheet. This analysis provided insight into the stretching of the cell flow field. Despite the wide variety of motion within the sheet, we found that its flow is not chaotic, since all values of the FTLE were less than zero. Although we do expect stretching within the monolayer, as our assay is based on a radially expanding sheet of cells, our FTLE measurements indicate that the stretching is not exponential in time. Near the very edge of the migrating sheet is the most common location where we see positive FTLE values; thus, the most rapid radial stretching is found in a small region close to the leading edge. We also find that increased stimulation through higher concentrations of serum leads to more instances of chaotic motion. Although the FTLE metric does not provide insight into the migration mechanism, these results suggest that it could be of use in quantifying future studies on more specific sources of stimulation.

Approaching the migrating sheet as a collection of particles instead of a flowing continuum allowed us to address another aspect of the epithelial sheet motion: rearrangement. Using $D^2_{min}$ as a parameterization of the non-affine motion within the monolayer, we found an increase in rearrangements when moving from the leading edge of the sheet to the center. Using non-affine motion as a metric for rearrangements provides more specific information than other metrics such as pair separation distance (see for example the work by Ng et al [37]). Pair separation distance describes the motion of two cells away from each other over time; although this could be a sign of cell mixing, it could also signal motions such as smooth spreading. Considering $D^2_{min}$ as a measure of non-affine motion allows us to separate smooth spreading from actual cell rearrangements. It also allows to retain information about the smooth motion of the sheet (such as spreading or translation) to compare to the non-affine individualistic motion seen in individual cells.

Looking only at the variation of rearrangements in comparison to radial position within the monolayer, we saw the same variation in properties seen when looking at simpler metrics such as speed. Motivated by similar trends seen in local cell number density, we compared the rearrangements to the number of neighbors around each cell. We found a clear relationship between the number density and rearrangements despite the day-to-day variability seen in other metrics. While this motivated the study of non-affine motion within cell sheets, even more compelling is that an increase in local number density was accompanied by an increase in rearrangements. This trend opposes the trends seen in current research on inanimate objects [36], suggesting that it is a uniquely active property of the cell system.

Our analysis of the monolayer migration in this simple radial assay provides quantitative methods and metrics that could be of use in other migration assays. As migration assays are widely used in drug studies and in determination of metastatic potential, these metrics could be readily introduced in the study of cancer to investigate differences in migration between cell lines of differing metastatic potential. Quantitative measurement of stretching and rearrangement could also be used to distinguish between the effects of various treatments designed to combat aggressive tumor cells. As migration studies become increasingly complex in an attempt to combat the wide variety of phenotypes seen in cancer and other diseases, quantitative measures such as these will become increasingly important.

**Acknowledgments**

We thank Michael Weiger, Christina Stuelten, and Carole Parent for their advice on cell culture and the use of their laboratory. This work was funded in part by the Intramural Research Program of the Center for Cancer Research, NCI, National Institutes of Health. The work was also supported by a National Science Foundation Graduate Research Fellowship to Rachel Lee. Wolfgang Losert was supported by a grant from the National Institutes of Health (NIH-R01GM085574).

**Figures**

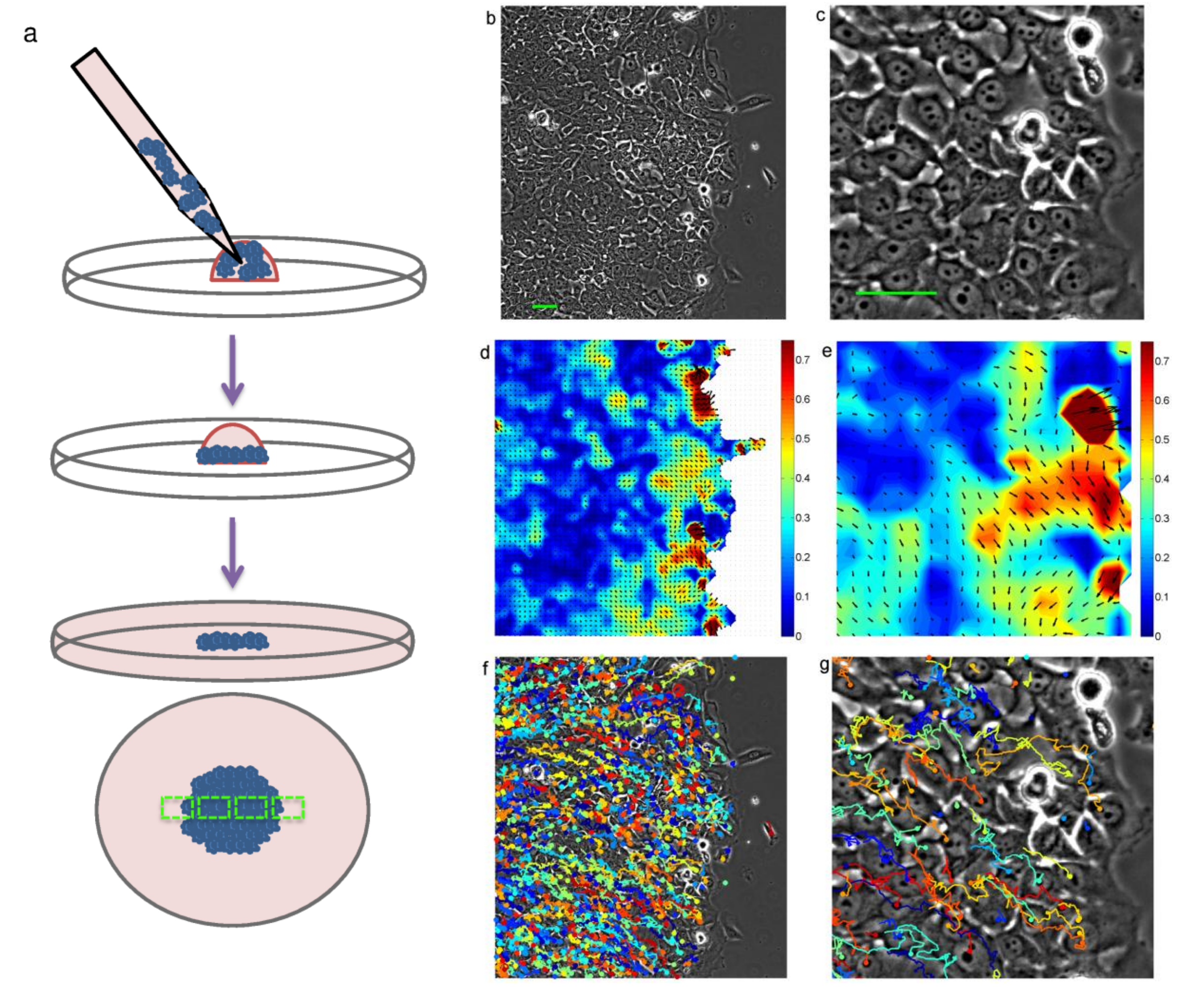

**Figure 1**

(a) Cartoon of migration assay. Cells were plated in a 2 μl drop and allowed to adhere the collagen surface before the well was filled with media. (b,c) Phase contrast image of leading edge of cell migration (scale bar, 50 μm). (d,e) Velocity field from PIV analysis. The color map represents the speed (μm/min) and is overlaid with velocity vectors. (f,g) Example of particle tracking overlaid on phase contrast image. Current particle positions are marked with a circle and the previous positions of the cell are marked by a track of the same color. (c,e,g) Enlargements of (b,d,f).

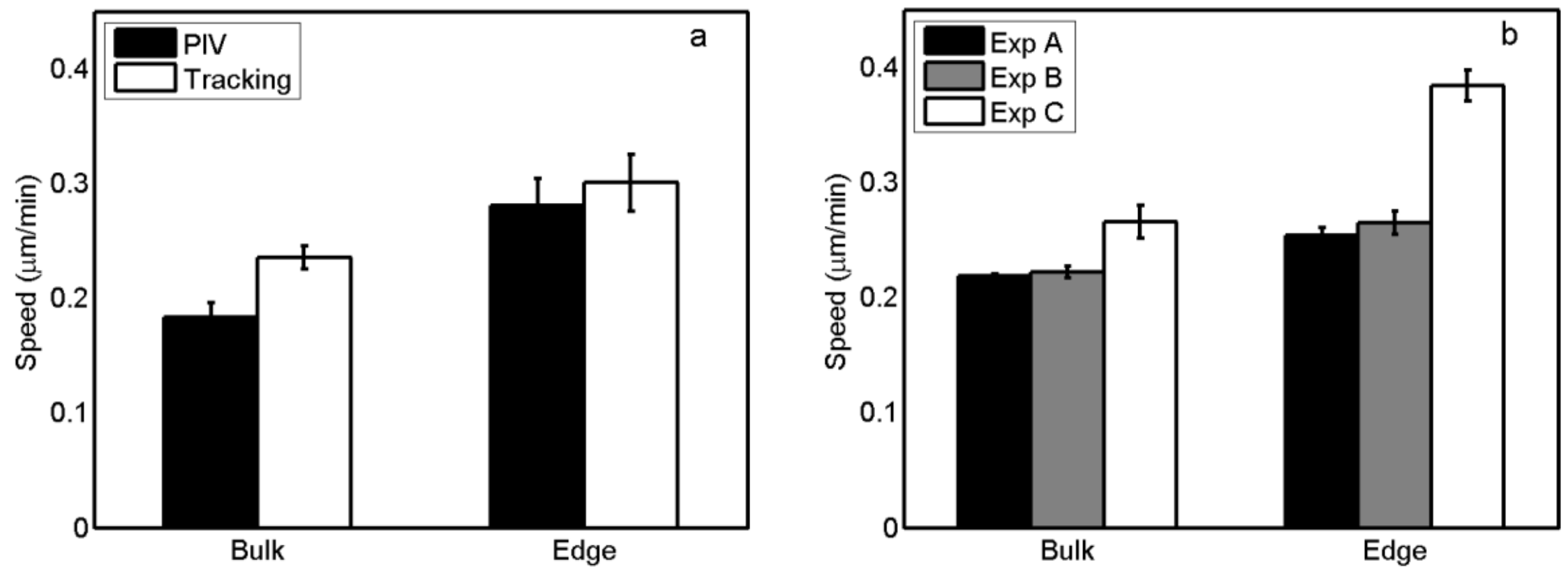

**Figure 2**

(a) A comparison of the mean speeds between the bulk and edge movies highlighting the distinction between PIV and nuclear tracking (composite data from three experiments) (b) A comparison of the means speeds from tracking highlighting the variability between experiments.

Error bars represent 95% confidence interval.

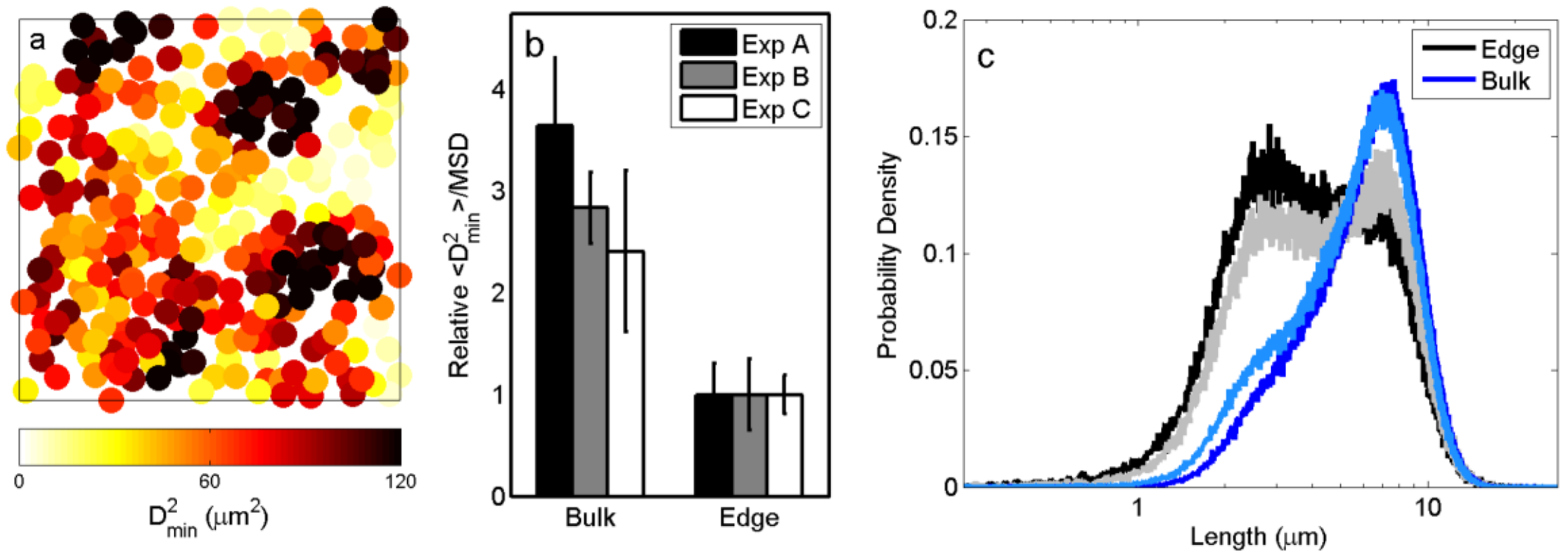

**Figure 3**

(a) A snapshot of individually tracked cells in the bulk, with each cell's value of $D^2_{min}$ represented by the color. Regions of high rearrangement ("hot spots") are seen. (b) A comparison of mean $D^2_{min}$ values between edge and bulk, with values normalized by the mean square displacement. The variability between experiments is also shown. Error bars represent 95% confidence interval. (c) A probability distribution of $D^2_{min}$ values for two representative edge movies and two corresponding bulk movies. The relative weight placed on the two different peaks suggests a difference in the modes of rearrangement seen at the edge and in the bulk.

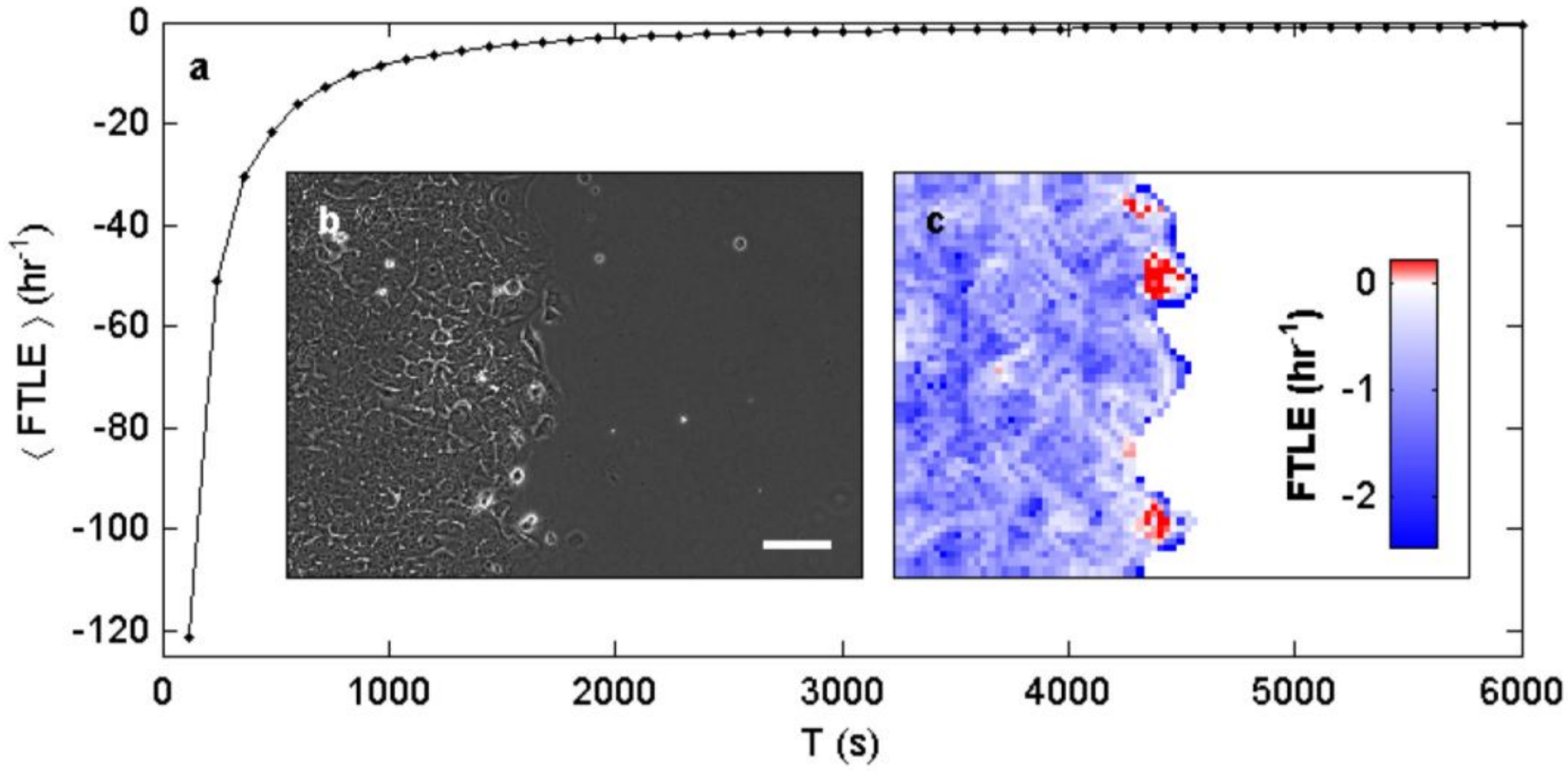

**Figure 4**

(a) Spatial mean of the FTLE field, varying with deformation time *T*. (b), A single phase-contrast image, along with (c) the corresponding FTLE field. The scale bar is 100 μm.

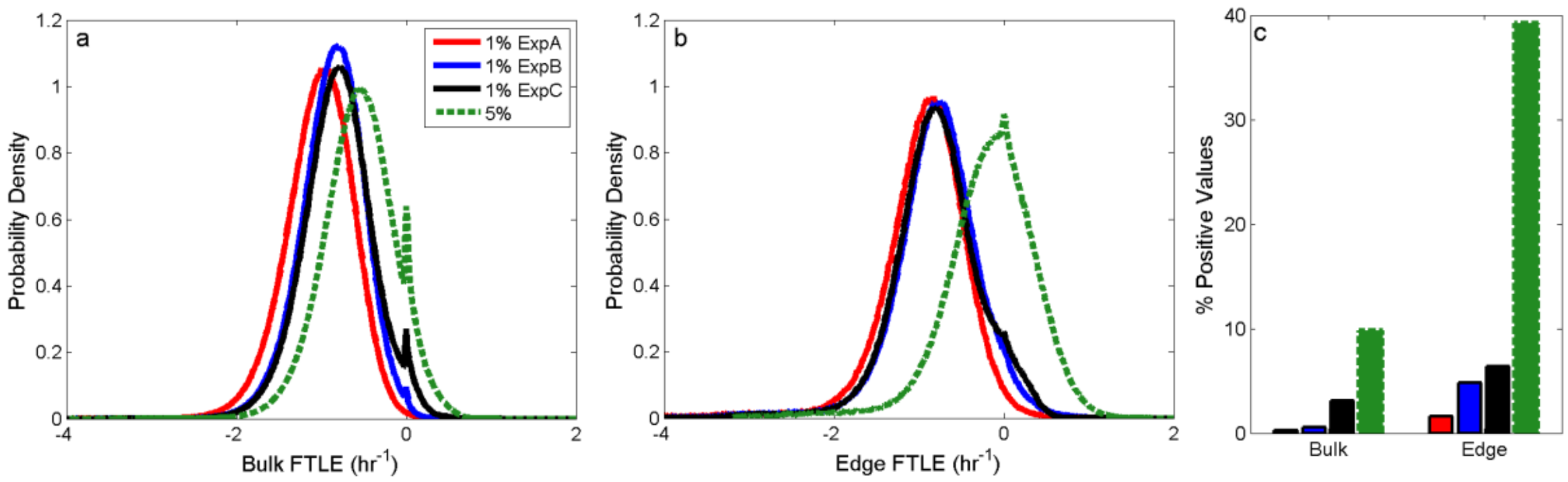

**Figure 5**

Probability densities of FTLE fields for measurements in the bulk (a) and at the edge (b), with multiple solid lines representing average results from three different days of experiments with 1% horse serum compared to the more chaotic distributions seen in a set of migration experiments at 5% horse serum (dashed line). (c) Fraction of FTLE values which were positive (i.e. chaotic) for four different days of experiments. A higher concentration of horse serum corresponds to an increased instance of chaotic regions.

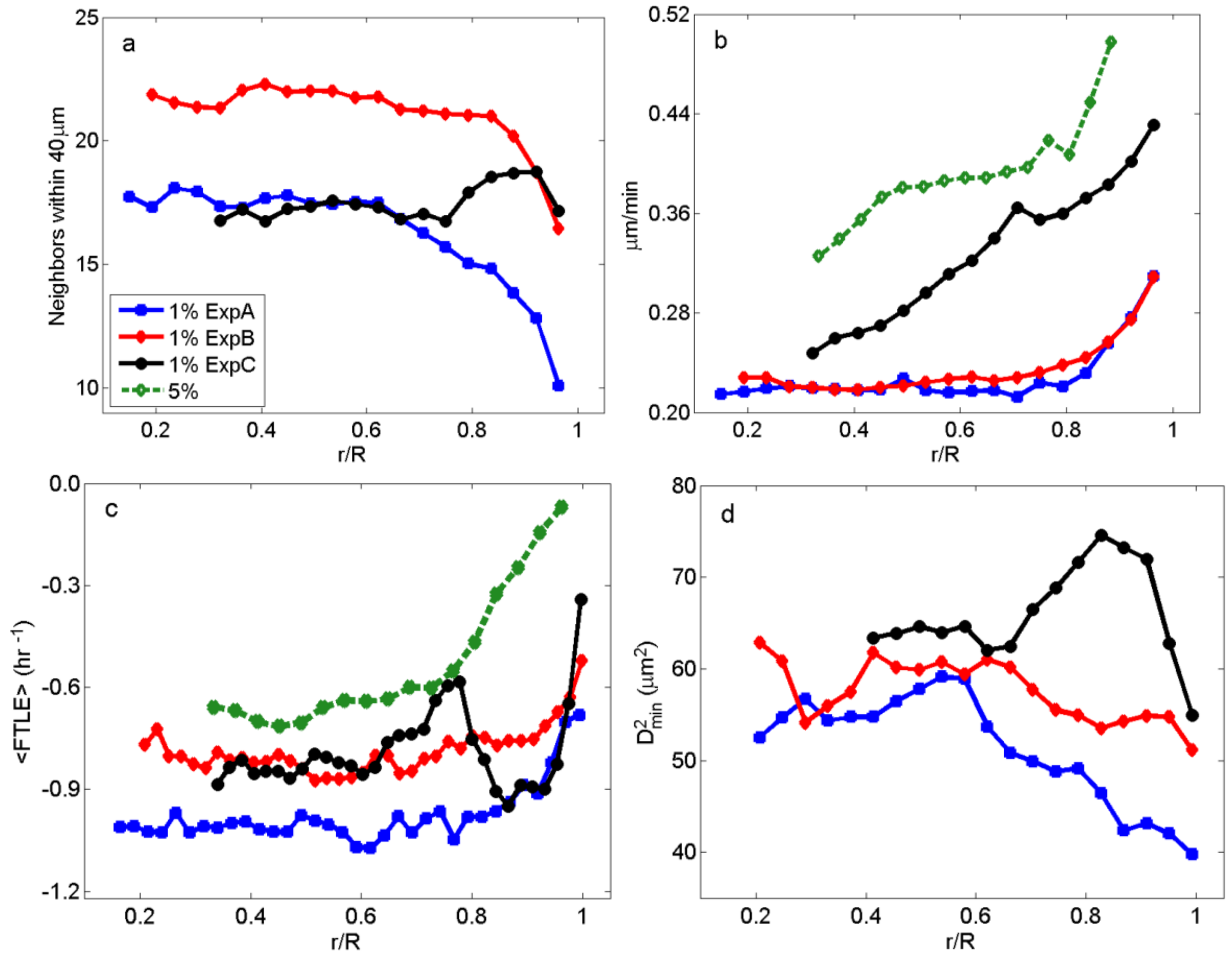

**Figure 6**

(a) Number density versus the normalized radius (*r/R*). In two cases, the density gradually decreases near the leading edge. In Exp C, a higher bump in density is seen near the leading edge. This is accompanied by a faster migration as seen in the speed profiles in (b). (c) FTLEs have a negative mean that approaches zero at long deformation times, showing that the flow is not chaotic. But FTLEs are typically higher near advancing edges, where stretching is large. Higher FTLE values were observed when the cells were stimulated by higher concentrations of serum (5% serum compared to 1% serum). (d) $D^2_{min}$ is in general higher near the center of the dot, with fewer rearrangements seen near the leading edge. In the case of the Exp C, a spike in $D^2_{min}$ is seen which corresponds to the same region of higher density seen in (a).

Each line represents the average over four different migration assays performed on the same day.

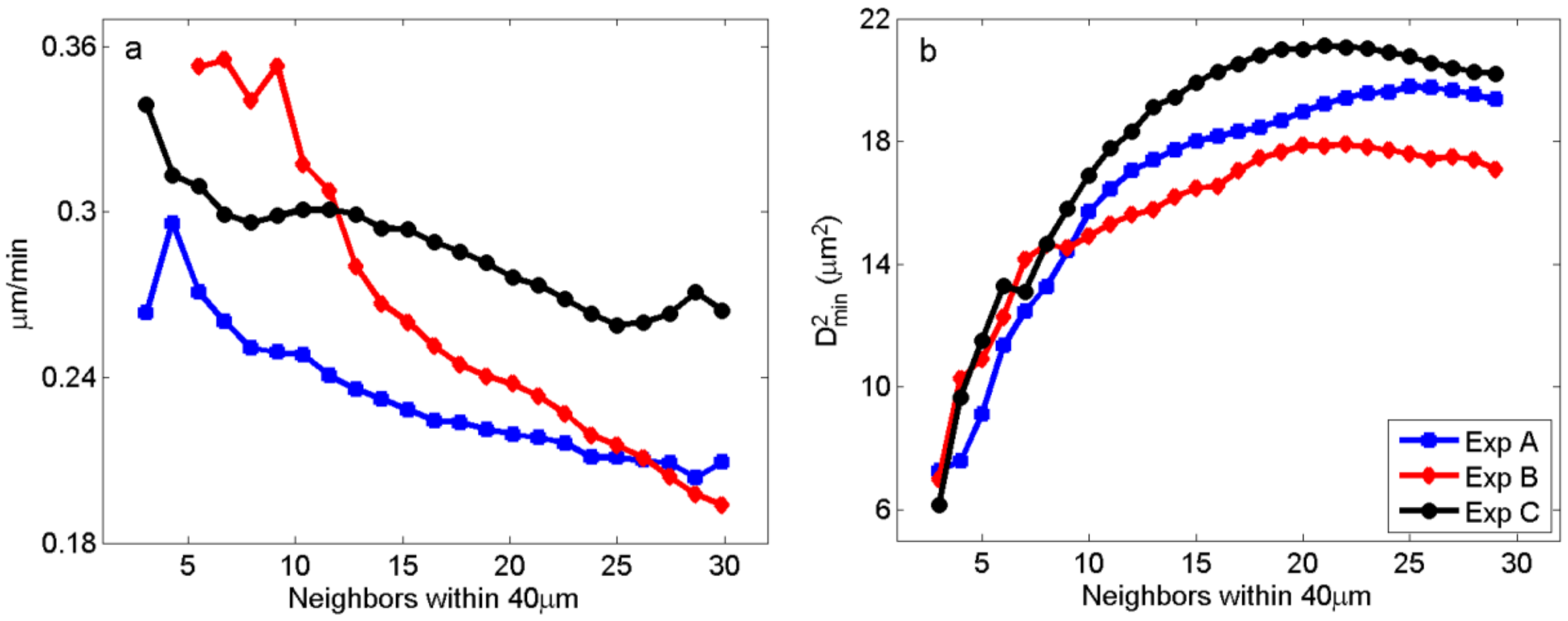

**Figure 7**

Speed and rearrangements can depend on cell density, even in the bulk movies where there is no free edge. (a) Although the magnitude of the speed is dependent on the variation between experiments, in all cases, speed decreases at higher cell densities. (b) $D^2_{min}$ shows a clear dependence on density, with rearrangements increasing as number density increases.

Each line is calculated from one representative bulk movie from each day of experiments.

**References**

[1] Weijer C J 2009 Collective cell migration in development. *Journal of cell science* **122** 3215–23

[2] Blanchard G B, Murugesu S, Adams R J, Martinez-Arias A and Gorfinkiel N 2010 Cytoskeletal dynamics and supracellular organisation of cell shape fluctuations during dorsal closure. *Development* **137** 2743–52

[3] Poujade M, Grasland-Mongrain E, Hertzog A, Jouanneau J, Chavrier P, Ladoux B, Buguin A and Silberzan P 2007 Collective migration of an epithelial monolayer in response to a model wound *Proceedings of the National Academy of Sciences* **104** 15988–93

[4] Mark S, Shlomovitz R, Gov N S, Poujade M, Grasland-Mongrain E and Silberzan P 2010 Physical model of the dynamic instability in an expanding cell culture. *Biophysical journal* **98** 361–70

[5] Vitorino P, Hammer M, Kim J and Meyer T 2011 A steering model of endothelial sheet migration recapitulates monolayer integrity and directed collective migration. *Molecular and cellular biology* **31** 342–50

[6] Indra I, Undyala V, Kandow C, Thirumurthi U, Dembo M and Beningo K A 2011 An in vitro correlation of mechanical forces and metastatic capacity. *Physical Biology* **8** 015015

[7] Kumar S and Weaver V M 2009 Mechanics, malignancy, and metastasis: the force journey of a tumor cell. *Cancer Metastasis Reviews* **28** 113–27

[8] Friedl P and Wolf K 2003 Tumour-cell invasion and migration: diversity and escape mechanisms. *Nature Reviews Cancer* **3** 362–74

[9] Szabó A, Varga K, Garay T, Hegedűs B and Czirók A 2012 Invasion from a cell aggregate-the roles of active cell motion and mechanical equilibrium. *Physical biology* **9** 016010

[10] Trepat X and Fredberg J J 2011 Plithotaxis and emergent dynamics in collective cellular migration *Trends in cell biology* **21** 638–46

[11] du Roure O, Saez A, Buguin A, Austin R H, Chavrier P, Silberzan P and Ladoux B 2005 Force mapping in epithelial cell migration. *Proceedings of the National Academy of Sciences* **102** 2390–5

[12] Saez A, Anon E, Ghibaudo M, du Roure O, Di Meglio J-M, Hersen P, Silberzan P, Buguin A and Ladoux B 2010 Traction forces exerted by epithelial cell sheets. *Journal of Physics: Condensed matter* **22** 194119

[13] Trepat X, Wasserman M R, Angelini T E, Millet E, Weitz D a., Butler J P and Fredberg J J 2009 Physical forces during collective cell migration *Nature Physics* **5** 426–30

[14] Tambe D T, Hardin C C, Angelini T E, Rajendran K, Park C Y, Serra-Picamal X, Zhou E H, Zaman M H, Butler J P, Weitz D a., Fredberg J J and Trepat X 2011 Collective cell guidance by cooperative intercellular forces. *Nature materials* **10** 469–75

[15] Omelchenko T and Hall A 2012 Myosin-IXA Regulates Collective Epithelial Cell Migration by Targeting RhoGAP Activity to Cell-Cell Junctions. *Current Biology* **22** 278–88

[16] Weber G F, Bjerke M A and DeSimone D W 2012 A Mechanoresponsive Cadherin-Keratin Complex Directs Polarized Protrusive Behavior and Collective Cell Migration *Developmental Cell* **22** 104–15

[17] Kim J-H, Dooling L J and Asthagiri A R 2010 Intercellular mechanotransduction during multicellular morphodynamics. *Journal of the Royal Society, Interface* **7** S341–S350

[18] Murrell M, Kamm R and Matsudaira P 2011 Substrate viscosity enhances correlation in epithelial sheet movement *Biophysical journal* **101** 297–306

[19] Puliafito A, Hufnagel L, Neveu P, Streichan S J, Sigal A, Fygenson D K and Shraiman B I 2012 Collective and single cell behavior in epithelial contact inhibition *Proceedings of the National Academy of Sciences* **109** 739–44

[20] Nnetu K D, Knorr M, Strehle D, Zink M and Kas J A 2012 Directed persistent motion maintains sheet integrity during multi-cellular spreading and migration *Soft Matter* **8** 6913–21

[21] Szabo B, Szollosi G, Gonci B, Juranyi Z, Selmeczi D and Vicsek T 2006 Phase transition in the collective migration of tissue cells: experiment and model *Physical Review E* **74** 061908

[22] Angelini T E, Hannezo E, Trepat X, Marquez M, Fredberg J J and Weitz D a. 2011 Glass-like dynamics of collective cell migration. *Proceedings of the National Academy of Sciences* **108** 4714–9

[23] Petitjean L, Reffay M, Grasland-Mongrain E, Poujade M, Ladoux B, Buguin A and Silberzan P 2010 Velocity fields in a collectively migrating epithelium. *Biophysical journal* **98** 1790–800

[24] Li K, Miller E D, Chen M, Kanade T, Weiss L E and Campbell P G 2008 Cell population tracking and lineage construction with spatiotemporal context. *Medical Image Analysis* **12** 546–66

[25] Shadden S, Lekien F and Marsden J 2005 Definition and properties of Lagrangian coherent structures from finite-time Lyapunov exponents in two-dimensional aperiodic flows *Physica D: Nonlinear Phenomena* **212** 271–304

[26] Voth G, Haller G and Gollub J 2002 Experimental Measurements of Stretching Fields in Fluid Mixing *Physical Review Letters* **88** 254501(4)

[27] Haller G and Yuan G 2000 Lagrangian coherent structures and mixing in two-dimensional turbulence *Physica D: Nonlinear Phenomena* **147** 352–70

[28] Chen D, Semwogerere D, Sato J, Breedveld V and Weeks E R 2010 Microscopic structural relaxation in a sheared supercooled colloidal liquid *Physical Review E* **81** 011403

[29] Falk M L and Langer J S 1998 Dynamics of viscoplastic deformation in amorphous solids *Physical Review E* **57** 7192–205

[30] Utter B and Behringer R 2008 Experimental Measures of Affine and Nonaffine Deformation in Granular Shear *Physical Review Letters* **100** 208302(4)

[31] Kondic L, Fang X, Losert W, O’Hern C S and Behringer R P 2012 Microstructure evolution during impact on granular matter *Physical Review E* **85** 011305

[32] Murdoch, N, Rozitis B, Nordstrom K, Green S, Michel P, De Lophem T-L, and Losert W 2013 Granular Convection in Microgravity *Physical Review Letters* **110** 018307

[33] Kelley D H and Ouellette N T 2011 Using particle tracking to measure flow instabilities in an undergraduate laboratory experiment *American Journal of Physics* **79** 267–73

[34] Ouellette N T, Xu H and Bodenschatz E 2006 A quantitative study of three-dimensional Lagrangian particle tracking algorithms *Experiments in Fluids* **40** 301–13

[35] Kelley D H and Ouellette N T 2011 Separating stretching from folding in fluid mixing *Nature Physics* **7** 477–80

[36] Ellenbroek W G, van Hecke M and van Saarloos W 2009 Jammed frictionless disks: Connecting local and global response *Physical Review E* **80** 061307

[37] Ng M R, Besser A, Danuser G, & Brugge J S 2012 Substrate stiffness regulates cadherin-dependent collective migration through myosin-II contractility *The Journal of cell biology*, **199**